# Tilted Two Fluids Cosmological Models with Variable G and Λ In General Relativity

**V. J. Dagwal and D. D. Pawar***

Dept. of Mathematics, Govt. College of Engineering, Nagpur- 441 108, (India)

*School of Mathematical Sciences, Swami Ramanand Teerth

Marathwada University, Vishnupuri Nanded-431 606 (India)

E-mail: vdagwal@gmail.com, dypawar@yahoo.com

**Abstract:**

Tilted two fluids cosmological models with variable G and Λ In General Relativity are presented. Here one fluid is matter field modelling material content of the universe and another fluid is radiation field modelling the cosmic microwave background (CMB). The tiltedness is also considered .To get the deterministic model; we have assumed a supplementary condition $G(t) = sT^n$ where s and n are constants. We have also discussed the behaviours of some physical parameters.

## 1. Introduction:

In Einstein field equations, Newtonian constant of gravitation G plays the role of a coupling constant between geometry of space and matter. A number of authors investigated cosmological model with variable Newtonian gravitational constant G and cosmological constant Λ. The Cosmological Constants was proposed by Dirac [1]. Bertolami [2] has examined a cosmological model with a time-dependent cosmological term. The cosmological constant problem obtained by Weinberg [3]. A model with a cosmological term of the form Λ = β (a) where a is the scale factor of the universe and β is a positive constant derived by Overdin and Cooperstock [4]. Dolgov and Silk [5] have presented some of the recent discussions on the cosmological constant 'problem' and consequence on cosmology with a time-varying cosmological constant. Berman [6] has studied Cosmological models with variable gravitational and cosmological constants. Einstein field equations that treated G and Λ as coupling variables within the framework of general relativity investigated by Abdel-

Rahman[7]. Flat FRW models with variable G and variable $\Lambda$ are derived by Kalligas et al. [8]. Pradhan and Chakrabarty [9] have examined LRS Bianchi I models with varying gravitational and cosmological constants. Early viscous universe with variable gravitational and cosmological constants are proposed by Singh et al. [10]. Bali and Tinker [11] have obtained Bianchi type III bulk viscous barotropic fluid cosmological models with variable G and $\Lambda$. Dagwal [12] has formulated Kaluza-Klein viscous fluid cosmological model with A time dependent $\Lambda$.

Biachi type-I two fluid cosmological models with a variable G and $\Lambda$ has obtained by oli [13]. Samanta [14] has formulated presented two fluid anisotropic cosmological model with variable G and $\Lambda$. Singh et al.[15] have investigated two-fluid cosmological model in Bianchi type V space time without variable G and $\Lambda$.

Many researchers have presented several aspects of two fluid cosmological models without variable G and $\Lambda$. Cosmological models with two fluids evaluated by McIntosh [16]. Coley and Dunn [17] have examined Bianchi type $VI_0$ model with two fluid sources. Two fluid Bianchi type II cosmological models are developed by Pant and Oli [18]. Verma [19] has obtained Qualitative analysis of two fluids FRW cosmological models. Two fluid cosmological models in Bianchi type V space-time constructed by Adhav et al. [20]. Pawar and Dagwal [21] have investigated Bianchi type IX two fluids cosmological models in General Relativity. Venkateswarlu [22] has studied Kaluza-Klein mesonic cosmological model with two-fluid source. Two-fluid cosmological model of Bianchi type-V with negative constant deceleration parameter are investigated by Singh et al. [23]. Axially Bianchi type-I Mesonic cosmological models with two fluid sources in Lyra Geometry and Two fluid Axially Symmetric Cosmological Models in $f(R,T)$ Theory of Gravitation presented by Pawar et al. [24, 25].

In recent years, there has been a considerable interest in investigating spatially homogeneous and anisotropic cosmological models in which matter does not move orthogonal to the hyper surface of homogeneity. Such types of models are called tilted cosmological models. The general dynamics of tilted cosmological models are presented by King and Ellis [26]; Ellis and King [27]; Collins and Ellis [28]. Dunn and Tupper [29] have studied tilted Bianchi type-I cosmological model for perfect fluid. Tilted electromagnetic Bianchi type-I cosmological model in General Relativity calculated by Lorentz [30]. Mukherjee [31] has examined Bianchi type-I cosmological model with heat flux in General

Relativity. Different aspects of tilted cosmological models examined by Lidsey [32], Horwood et al [33], Bogoyavlenskii and Novikov [34], Hewitt et al. [35, 36], and Apostolopoulos [37]. Bali and Meena [38] have investigated tilted cosmological models filled with disordered radiation in General Relativity. Tilted plane symmetric cosmological models with heat conduction and disordered radiation studied by Pawar et al. [39]. Conformally flat tilted cosmological models and tilted Kantowski-Sachs cosmological models with disordered radiation in scalar tensor theory of gravitation proposed by Saez and Ballester are calculated by Pawar and Dagwal [40, 41]. Tilted Bianchi type VI0 cosmological model in Saez and Ballester scalar tensor theory of gravitation and Bianchi type-I mesonic stiff fluid cosmological model formulated by Sahu [42, 43]. Pawar and Dagwal [44 - 46] have obtained tilted plane symmetric magnetized cosmological models, tilted Cosmological Models in $f$(R,T) Theory of Gravitation and tilted Kasner-Type Cosmological Models in Brans-Dicke Theory of Gravitation.

Coley and Hervik [47] have presented Bianchi cosmologies a Tale of two tilted fluids. Bianchi type-I models with two tilted fluids derived by Sandin and Uggla [48]. Sandin [49] has constructed tilted two fluid Bianchi type-I models. Two fluids tilted cosmological model in General Relativity presented by Pawar and Dagwal [50]. Tilted and non tilted homogeneous plane symmetric C-field cosmological models are investigated by Pawar et al. [51]

Different aspects LRS Bianchi type-I space time presented by Thorne [52], Tripathy et al. [53, 54], Bali and Kumawat [55], Abdussattar and prajapati [56]. Pawar et al. [57, 58] have obtained bulk viscous fluid with plane symmetric string dust magnetized cosmological model in general relativity and Lyra manifold. Bayaskar et al. [59] have derived cosmological models of perfect fluid and massless scalar field with electromagnetic field.

The solution of the field equation can be established by applying a law of variation for Hubble' parameter which was investigated by Berman [60], that yields a constant value of deceleration parameter. The cosmological models with a constant deceleration parameter developed by Berman and Gmide [61]; Beesham [62]; Reddy and Venkateswara Rao [63]; pradhan and Vishwakarma [64]; Tiwari [65]. Two fluids Kantowski-Sachs cosmological models with matter and radiating source in scalar tensor theory of gravitation proposed by Saez and Ballester are studied by Pawar et al. [66, 67].

## 2. Field Equation:

We consider the metric in the form

$$ds^2 = dt^2 - R^2 \left\{ dx^2 + dy^2 + \left( 1 + \beta \int \frac{dt}{R^3} \right)^2 dz^2 \right\},$$  (1)

where R is functions of $t$ alone.

The Einstein's field equations are

$$R_j^i - \frac{1}{2} g_j^i R = -G(t) T_j^i - \wedge(t) g_j^i.$$  (2)

The energy momentum tensor for a two fluid source is given by

$$T_{ij} = T_{ij}^{(m)} + T_{ij}^{(r)},$$  (3)

where $T_{ij}^{(m)}$ is the energy momentum tensor for matter field and $T_{ij}^{(r)}$ is the energy momentum tensor for radiation field which are given by

$$T_{ij}^{(m)} = \left( p_m + \rho_m \right) u_i u_j - p_m g_{ij} + q_i u_j + q_j u_i,$$  (4)

$$T_{ij}^{(r)} = \frac{4}{3} \rho_r v_i v_j - \frac{1}{3} \rho_r g_{ij}$$  (5)

with   $g^{ij} u_i u_j = 1,$  (6)

$$q_i q^i > 0, \quad q_i u^j = 0,$$  (7)

and

$$g^{ij} v_i v_j = 1.$$  (8)

Where $p_m$ is the pressure, $\rho_m$ is the energy density for matter field and $\rho_r$ is the energy density for radiation field, $q_i$ is the heat conduction vector orthogonal to $u^i$. The fluid vector $u_i$ has the components $\left( R \sinh \alpha, \, 0, \, 0, \, \cosh \alpha \right)$ for matter field satisfying Equation (6) and

$\alpha$ is the tilt angle. The fluid vector $v_i$ has the components (0, 0, 0, 1) for radiation field satisfying Equation (8)

The Einstein's field equation (2) reduces to

$$\frac{R_4^2}{R^2} + 2\frac{R_{44}}{R} = -G(t)\left[\left(\rho_m + P_m\right)\sinh^2\alpha + p_m + \frac{\rho_r}{3} + 2q_1\frac{\sinh\alpha}{R}\right] + \wedge(t) \quad,\tag{9}$$

$$\frac{R_4^2}{R^2} + 2\frac{R_{44}}{R} = -G(t)\left[p_m + \frac{\rho_r}{3}\right] + \wedge(t) \quad,\tag{10}$$

$$3\frac{R_4^2}{R^2} + \frac{2\beta R_4}{R^4\left(1 + \beta\int\frac{dt}{R^3}\right)} = G(t)\left[\left(\rho_m + P_m\right)\cosh^2\alpha - p_m + \rho_r + 2q_1\frac{\sinh\alpha}{R}\right] + \wedge(t) \quad,\tag{11}$$

$$G(t)\left[\left(\rho_m + P_m\right)R\sinh\alpha\cosh\alpha + q_1\cosh\alpha + q_1\frac{\sinh^2\alpha}{\cosh\alpha}\right] = 0 \quad.\tag{12}$$

We solve the above set of highly non linear equations with the help of special law of variation of Hubble's parameter, proposed by Berman (1983) that yields constant deceleration parameter model of universe. We consider only constant deceleration parameter model defined by

$$q = -\left[\frac{A A_{44}}{A_4^2}\right],\tag{13}$$

where $A = \left[R^3\left(1 + \beta\int\frac{dt}{R^3}\right)\right]^{1/3}$ is the overall scalar factor.  (14)

Here the constant is taken as negative because the sign of deceleration parameter q whether the model accelerates or not .The positive sign q (>1) corresponds the decelerating model where the negative sign (-1 < q < 0) indicates acceleration and q = 0 corresponds to expansion with constant velocity.  The solution  of  (13) is given by

$$A = \left(Ct + D\right)^{1/{1+q}} \quad,\tag{15}$$

where C & D are constant of integration.

This equation implies that the condition of expansion is $1+q > 0$.

From equations (14) and (15), we get

$$\left[ R^3 \left( 1 + \beta \int \frac{dt}{R^3} \right) \right]^{1/3} = (Ct + D)^{1/1+q} .$$

(16)

Solving equation (16) we get

$$R = T^{1/1+q} \, e^{NT^{\frac{q-2}{1+q}}} ,$$

(17)

where $(Ct + D) = T$ & $N = \dfrac{\beta (1+q)}{3C (2-q)}$.

The metric (1) reduces to the following form

$$ds^2 = dT^2 - T^{2/1+q} e^{2NT^{\frac{q-2}{1+q}}} \left( dx^2 + dy^2 \right) - T^{2/1+q} e^{-4NT^{\frac{q-2}{1+q}}} dz^2 ,$$

(18)

Where $(Ct + D) = T$ , $N = \dfrac{\beta (1+q)}{3C (2-q)}$ & $C = 1$.

## 3. Some Physical and Geometrical Properties

For the equation of state of matter we shall assume the gamma law

$$p_m = (\gamma - 1)\rho_m , \qquad 1 \le \gamma \le 2 .$$

(19)

We assume a power law of gravitational constant G

$$G(t) = sT^n ,$$

(20)

where s & n are constant.

Conservation law for radiation field

$$\rho_{r_4} + \left[ 4 \frac{R_4}{R} + \frac{4\beta}{3R^3 \left( 1 + \beta \int \frac{dt}{R^3} \right)} \right] \rho_r = 0 .$$

(21)

Solving equation (9), (10), (11), (17),(20) and (21) We get energy density of radiation, energy density matter and pressure for matter as

$$\rho_r = \frac{k}{T^{\frac{4}{1+q}}} \quad , \tag{22}$$

$$\rho_m = \frac{1}{\gamma}\left[\frac{1}{s}\left(\frac{2}{(1+q)T^{n+2}} - \frac{2\beta^2}{3T^{\frac{n(1+q)+6}{1+q}}}\right) - \frac{4k}{3T^{\frac{4}{1+q}}}\right], \tag{23}$$

and

$$p_m = \frac{(\gamma-1)}{\gamma}\left[\frac{1}{s}\left(\frac{2}{(1+q)T^{n+2}} - \frac{2\beta^2}{3T^{\frac{n(1+q)+6}{1+q}}}\right) - \frac{4k}{3T^{\frac{4}{1+q}}}\right], \tag{24}$$

where k is integration constant.

Solving equation (9), (10) (11), (22) & (24) the cosmological constant is given by

$$\wedge(t) = \left[\frac{3\gamma - 2(1+q)}{\gamma(1+q)}\right]\frac{1}{(1+q)T^2} + \left[\frac{2-\gamma}{\gamma}\right]\frac{\beta^2}{3T^{\frac{6}{1+q}}} + \frac{k\,s\left(4-3\gamma\right)}{3\gamma\,T^{\frac{4-n(1+q)}{1+q}}} \tag{25}$$

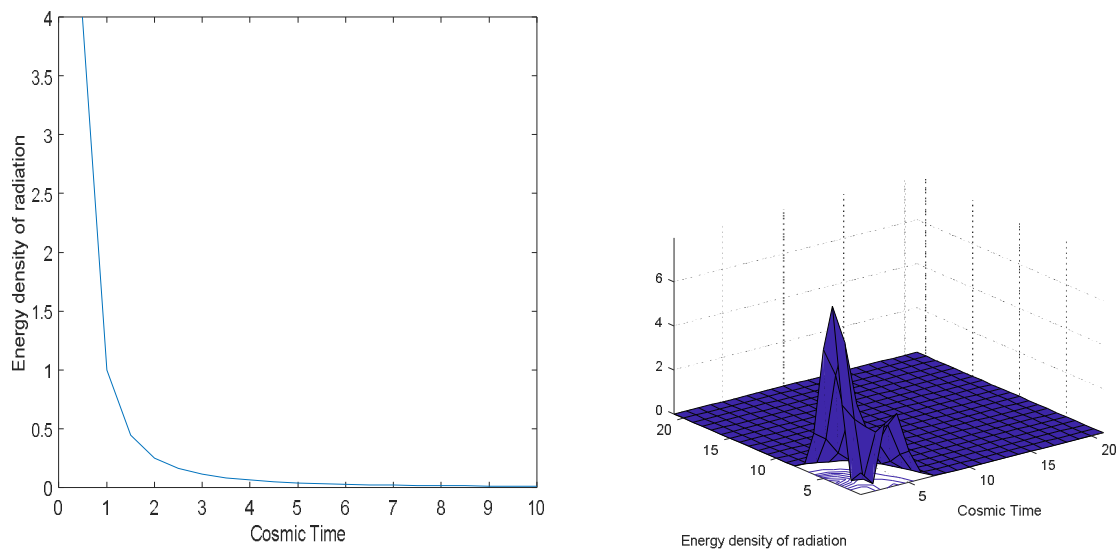

Fig. 1 Energy density of radiation versus Cosmic Time

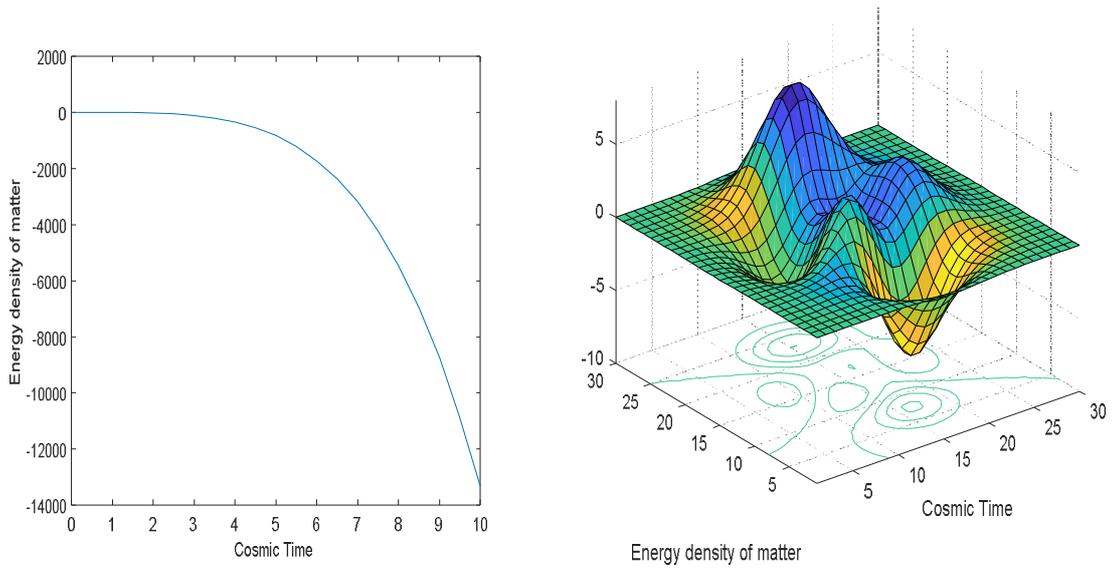

Fig.2 Energy density of matter versus Cosmic Time

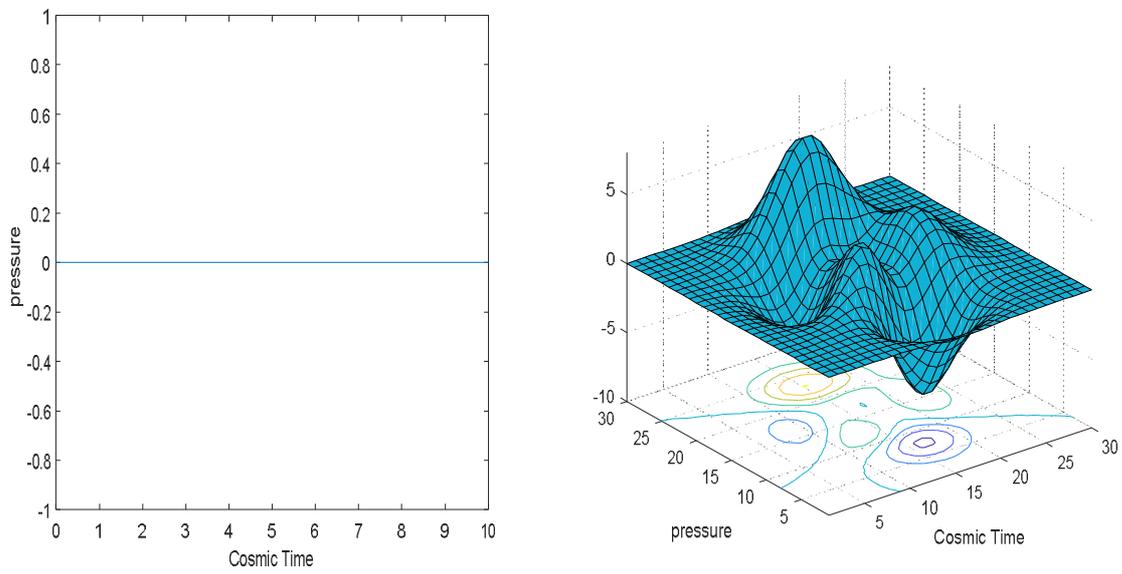

Fig.3 Pressure for matter versus Cosmic Time

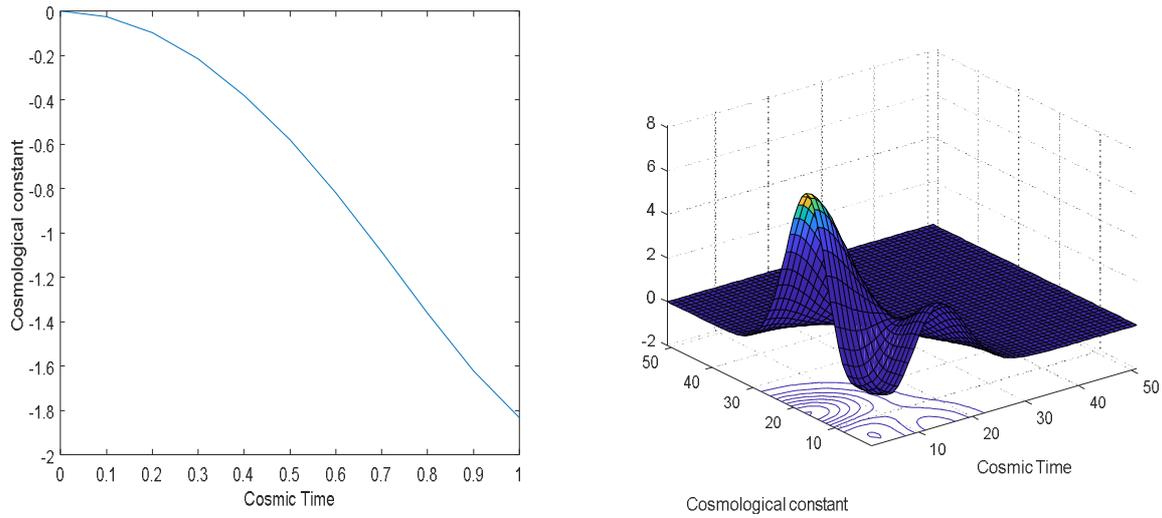

Fig.4 Cosmological constant versus Cosmic Time

In Fig.1 we show energy density for radiation versus cosmic time by setting the value $q=1$ & $k=1$. In Fig.2 we show energy density for matter versus cosmic time by setting the value $q=1$, $\gamma=1$, $s=1$, $n=1$, $\beta=1$ & $k=1$. In Fig.3 we show pressure for matter versus cosmic time by setting the value $q=1$, $\gamma=1$, $s=1$, $n=1$, $\beta=1$ & $k=1$ also in Fig.4 we show cosmological constant versus cosmic time by setting the value $q=1$, $\gamma=1$, $s=1$, $n=1$, $\beta=1$ & $k=1$. Initially, the energy density for radiation $\rho_r$ is infinite, the energy density for matter $\rho_m$ and the cosmological constant $\wedge$ are undetermined but for large value of T, the energy density for radiation $\rho_r$ and matter $\rho_m$, the cosmological constant $\wedge$ are vanish. For dust universe $\gamma=1$, the pressure $p_m$ is zero, the cosmological constant $\wedge$ is infinite at $q=-1$ and when $T=\infty$, the cosmological constant $\wedge$ is vanishing. At $T=\infty$ & $n=\dfrac{4}{1+q}$, the cosmological constant $\wedge$ is constant but the cosmological constant $\wedge$ is infinite, when $T=0$ & $n=\dfrac{4}{1+q}$ for Zeldovich universe $\gamma=2$. For radiation universe $\gamma=\frac{4}{3}$, the cosmological constant $\wedge$ is infinite at $T=0$ & $q=2$ but for large value of T and $q=2$, the cosmological constant $\wedge$ is zero.

In case $\beta=0$ & $k=0$

When the deceleration parameter $q = 1$, the cosmological constant $\wedge$ is zero for dust universe $\gamma = 1$. For Zeldovich universe $\gamma = 2$, the cosmological constant $\wedge$ is vanish at $q = 4$. When $q \to 2$, the cosmological constant $\wedge \to 0$ for radiation universe $\gamma = \frac{4}{3}$.

The tilt angle $\alpha$, flow vectors $u^i$ and heat conduction vectors $q_i$ for the model (18) are given by

$$\cosh\alpha = 1 \quad \& \quad \sinh\alpha = 0 \quad , \tag{26}$$

$$u_1 = 0 \quad \& \quad u_4 = 1 \quad , \tag{27}$$

$$q_1 = 0 \quad \& \quad q_4 = 0 \quad . \tag{28}$$

Hubble parameter and spatial volume are given by

$$H = \frac{6}{(1+q)T} \quad , \tag{29}$$

$$V = T^{\frac{1}{1+q}} . \tag{30}$$

When $T = 0$, the Hubble parameter is infinite and spatial volume is zero but at $T = \infty$, the Hubble parameter is zero and spatial volume is infinite. For $q = -1$, the Hubble parameter and spatial volume are infinite.

The anisotropy parameter, scalar of expansion and shear scalar as

$$\Delta = \frac{2}{9}\left[ 2 + \frac{\beta^2(1+q)^2}{9T^{\frac{2(2-q)}{1+q}}} \right] , \tag{31}$$

$$\theta = \frac{6}{(1+q)T} \quad , \tag{32}$$

$$\sigma^2 = \frac{2\beta^2}{3T^{6/1+q}} . \tag{33}$$

For large value of T, the anisotropy parameter is constant, the scalar of expansion and shear scalar are vanish but initially, the anisotropy parameter , the scalar of expansion and shear scalar are infinite. At $q = 2$, the anisotropy parameter is constant .The scalar of expansion is infinite, the shear scalar is zero and the anisotropy parameter is constant for $q = -1$.When $\beta \to 0$, the model is nonshearing universe and the anisotropy parameter $\Delta \to \dfrac{4}{9}$.

The density parameters are

$$\Omega_m = \frac{(1+q)}{\gamma s}\left[\frac{1}{54T^n} - \frac{\beta^2(1+q)}{162T^{\frac{(n+4)+(n-2)q}{1+q}}} - \frac{k(1+q)}{81T^{\frac{2(1-q)}{1+q}}}\right],$$  (34)

$$\Omega_r = \frac{k(1+q)^2}{108T^{\frac{2(1-q)}{1+q}}}.$$  (35)

Total energy density as

$$\Omega = \frac{(1+q)}{\gamma s}\left[\frac{1}{54T^n} - \frac{\beta^2(1+q)}{162T^{\frac{(n+4)+(n-2)q}{1+q}}}\right] + \left[\frac{1}{108} - \frac{1}{81\gamma s}\right]\frac{k(1+q)^2}{T^{\frac{2(1-q)}{1+q}}}.$$  (36)

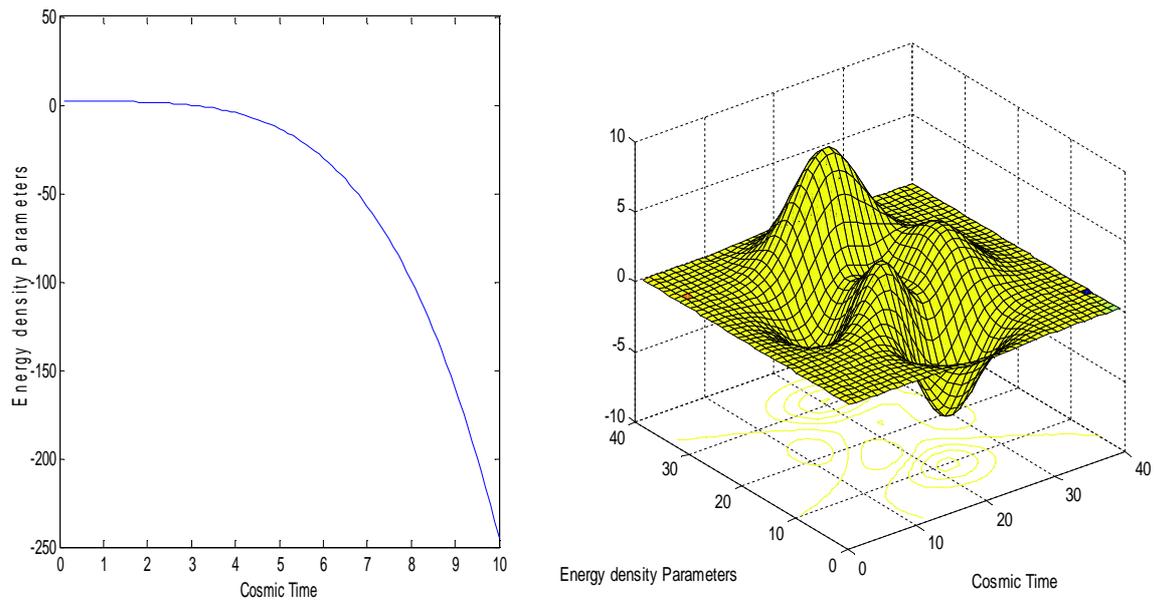

Fig.5 Density parameter versus Cosmic Time

In Fig.5 we show density parameter versus cosmic time by setting the value $q = 1$, $\gamma = 1$, $s = 1$, $n = 1$, $\beta = 1$ & $k = 1$. When $T \rightarrow \infty$, all density parameters are vanishing. The density parameter $\Omega_r$ is infinite, the density parameter $\Omega_m$ and total density parameter $\Omega$ are undetermined at $T \rightarrow 0$. All density parameters are vanishing for $q = -1$. At $s = 0$, the density parameter $\Omega_m$ and total density parameter $\Omega$ are infinite. The density parameter $\Omega_r$ is zero, when $k = 0$.

## 4. Conclusion:

We have presented tilted two fluids cosmological models with variable G and $\Lambda$ In General Relativity. The study results into an expanding and shearing universe. The tilt angle $\alpha$, flow vectors $u^i$ are constant and heat conduction vectors $q_i$ are zero. Initially, the energy density for radiation $\rho_r$ is infinite, the energy density for matter $\rho_m$ and the cosmological constant $\wedge$ are undetermined but for large value of T, the energy density for radiation $\rho_r$ and matter $\rho_m$, the cosmological constant $\wedge$ are vanish.

For dust universe $\gamma = 1$, the pressure $p_m$ is zero, the cosmological constant $\wedge$ is infinite at $q = -1$ and when $T = \infty$, the cosmological constant $\wedge$ is vanishing. At $T = \infty$ & $n = \dfrac{4}{1+q}$, the cosmological constant $\wedge$ is constant but the cosmological constant $\wedge$ is infinite, when $T = 0$ & $n = \dfrac{4}{1+q}$ for Zeldovich universe $\gamma = 2$. For radiation universe $\gamma = \dfrac{4}{3}$, the cosmological constant $\wedge$ is infinite at $T = 0$ & $q = 2$ but for large value of T and $q = 2$, the cosmological constant $\wedge$ is zero.

When $T = 0$, the Hubble parameter is infinite and spatial volume is zero but at $T = \infty$, the Hubble parameter is zero and spatial volume is infinite. For $q = -1$, the Hubble parameter and spatial volume are infinite. For large value of T, the anisotropy parameter is constant, the scalar of expansion and shear scalar are vanish but initially, the anisotropy parameter, the scalar of expansion and shear scalar are infinite. At $q = 2$, the anisotropy parameter is constant. The scalar of expansion is infinite, the shear scalar is zero and the anisotropy parameter is constant for $q = -1$. When $T \rightarrow \infty$, all density parameters are vanishing. The density parameter $\Omega_r$ is infinite, the density parameter $\Omega_m$ and total density parameter $\Omega$ are undetermined at $T \to 0$. All density parameters are vanishing for $q = -1$. At $s = 0$, the density parameter $\Omega_m$ and total density parameter $\Omega$ are infinite. The density parameter $\Omega_r$ is zero, when $k = 0$. When $\beta \to 0$, the model is nonshearing universe and the anisotropy parameter $\Delta \to \dfrac{4}{9}$.

In case $\beta = 0$ & $k = 0$

When the deceleration parameter $q = 1$, the cosmological constant $\wedge$ is zero for dust universe $\gamma = 1$. For Zeldovich universe $\gamma = 2$, the cosmological constant $\wedge$ is vanish at $q = 4$. When $q \to 2$, the cosmological constant $\wedge \to 0$ for radiation universe $\gamma = \frac{4}{3}$.

Since $\underset{T \to \infty}{\lim} \left( \dfrac{\sigma}{\theta} \right) = 0$ the models approach isotropy for large value of $T$.